\documentclass[twocolumn,epjc3]{svjour3}

\usepackage[T1]{fontenc}
\usepackage[utf8]{inputenc}

\usepackage{amsmath}
\usepackage{amssymb}
\usepackage{mathtools}
\usepackage{cuted}
\usepackage{graphicx}
\usepackage{microtype}

\journalname{Eur. Phys. J. C}

\title{Proton-PDF uncertainties in extracting nuclear PDFs from $W^\pm$ production in p+Pb collisions}

\author{Kari~J.~Eskola$^{1,2,}$\thanksref{e1},\,Petja~Paakkinen$^{1,2,3,}$\thanksref{e2},\,Hannu~Paukkunen$^{1,2,}$\thanksref{e3},\,Carlos~A.~Salgado$^{3,}$\thanksref{e4}}

\thankstext{e1}{e-mail: kari.eskola@jyu.fi}
\thankstext{e2}{e-mail: petja.k.m.paakkinen@jyu.fi}
\thankstext{e3}{e-mail: hannu.paukkunen@jyu.fi}
\thankstext{e4}{e-mail: carlos.salgado@usc.es}

\institute{University of Jyvaskyla, Department of Physics, P.O. Box 35, FI-40014 University of Jyvaskyla, Finland \and
Helsinki Institute of Physics, P.O. Box 64, FI-00014 University of Helsinki, Finland \and
Instituto Galego de F\'\i sica de Altas Enerx\'\i as (IGFAE), Universidade de Santiago de Compostela, E-15782 Galicia, Spain
}

\begin{document}

\maketitle

\abstract{
  We discuss the recent CMS Collaboration measurement of $W^\pm$ boson production in p+Pb collisions at 8.16 TeV in terms of the constraining power on nuclear parton distribution functions (PDFs). The impact of the free-proton PDF uncertainties on the nuclear PDF extraction is quantified by using a theoretical covariance-matrix method and Hessian PDF reweighting. We discuss different ways to mitigate these theoretical uncertainties, including self-normalization, forward-to-backward ratios and nuclear-modification ratios. It is found that none of these methods offer perfect cancellation of the free-proton PDFs but, with the present data uncertainties, the residual free-proton-PDF dependence has, conveniently for the global analyses, little effect on the extraction of the nuclear modifications. Based on a simple estimate of obtainable statistics at the LHC Run~3, we argue that this will change in the near future and it becomes more important to propagate the proton-PDF uncertainties accordingly. Using the obtained information on the correlations of the free-proton uncertainties, we also identify a new charge asymmetry ratio, where the cancellation of the proton-PDF uncertainties is found to be extremely good.
}

\section{Introduction}
\label{sec:intro}

The parton distribution functions (PDFs) of heavy nuclei, like their free-proton counterparts, are currently obtained most reliably from global analyses of experimental data. The bulk of these data comes from deep inelastic scattering (DIS) measurements which probe the nuclear structure directly and uniquely, with the precision limited at high enough scales only by experimental uncertainties and perturbative accuracy. However, to constrain the full flavour dependence of the nuclear PDFs, it is necessary to additionally use proton--nucleus (p+$A$) processes, including fixed-target Drell--Yan (DY) dilepton as well as collider electroweak (EW) boson and (di)jet production data. For these processes, the collinearly factorized cross sections contain a convolution of the nuclear and free-proton PDFs, and as a consequence, the nuclear PDFs extracted from such data become inherently dependent on the assumed free-proton PDFs.

One could then envisage two systematic approaches to treat the proton-PDF uncertainties in the nuclear-PDF analyses: First, one can try to reduce the proton-PDF uncertainties by using observables where one probes instead the \emph{nuclear modifications} of the PDFs, which are then parametrized and fitted, and the ``baseline'' free-proton PDF dependence effectively drops out, as has been done systematically in the EKS--EPPS line of analyses~\cite{Eskola:1998iy,Eskola:1998df,Eskola:2007my,Eskola:2008ca,Eskola:2009uj,Eskola:2016oht,Eskola:2021nhw}, and by others~\cite{Hirai:2001np,Hirai:2004wq,Hirai:2007sx,AtashbarTehrani:2012xh,Khanpour:2016pph}. This approach has been particularly attractive since much of the older DIS and DY data are in any case available only in terms of nuclear ratios. Or, second, one could allow using also absolute cross sections, taking into account all possible correlations with the free-proton PDFs, a program which has more recently been undertaken by the nNNPDF collaboration~\cite{AbdulKhalek:2019mzd,AbdulKhalek:2020yuc,Khalek:2022zqe}.

In some other instances the treatment of the baseline proton-PDF dependence have been less explicit~\cite{deFlorian:2003qf,deFlorian:2011fp,Walt:2019slu,Khanpour:2020zyu,Schienbein:2009kk,Kovarik:2015cma,Kusina:2020lyz,Segarra:2020gtj,Duwentaster:2021ioo,Helenius:2021tof}, with the inherent assumption being that the free-proton uncertainties are in any case smaller than the nuclear-PDF ones and thus do not cause a significant bias in the fit. Until very recently, this has been a justifiable approximation. However, as the precision of data from the LHC p+$A$ program improves, it can become necessary to either propagate or mitigate the proton-PDF uncertainties in extracting the nuclear PDFs.

One of the latest additions to the nuclear-PDF constraints is the CMS Collaboration measurement of $W^\pm$ boson production in LHC Run~2 p+Pb collisions at 8.16 TeV~\cite{CMS:2019leu}, with an eight-fold increase in the statistics compared to the Run~1 data taking at 5.02 TeV~\cite{CMS:2015ehw}. These data have been already included in nuclear-PDF analyses, where they have been seen to give constraints either specifically on the gluon and sea-quark PDFs~\cite{Helenius:2021tof}, strangeness~\cite{Kusina:2020lyz}, or on the flavour separation in more general~\cite{AbdulKhalek:2020yuc}. The level at which the proton-PDF uncertainties are treated in these analyses varies, with Refs.~\cite{Kusina:2020lyz,Helenius:2021tof} taking the proton-PDFs as fixed, ignoring their uncertainties, and Ref.~\cite{AbdulKhalek:2020yuc} propagating the proton-PDF uncertainties in the analysis, but not discussing their importance in the fit.

The role of the proton-PDF uncertainties in $W^\pm$ production in p+Pb collisions has been considered previously in Ref.~\cite{Paukkunen:2010qg}. In this paper, we elaborate their significance in the context of the aforementioned CMS Collaboration measurement~\cite{CMS:2019leu} (Section~\ref{sec:protonunc}) and different ratios constructed from the data (Section~\ref{sec:unc_red}). We also extend the analysis of the importance of the proton-PDF uncertainties in nuclear-modification fitting by the tools of theoretical covariance matrix (Section~\ref{sec:thcov}) and Hessian PDF reweighting (Section~\ref{sec:rw}).

\section{Proton-PDF uncertainties in $W^\pm$ production in p+Pb collisions}
\label{sec:protonunc}

It is conventional to write the PDFs $f_i^A$ of a nucleus with $Z$ protons and $N$ neutrons in terms of bound-nucleon PDFs at momentum fraction $x$ and scale $Q^2$ as
\begin{equation}
  f_i^A (x, Q^2) = Z f_i^{p/A} (x, Q^2) + N f_i^{n/A} (x, Q^2),
  \label{eq:isospin}
\end{equation}
taking the bound-neutron PDFs $f_i^{n/A}$ to be related to the bound-proton ones $f_i^{p/A}$ by the isospin symmetry:
\begin{equation}
  \begin{split}
    u^{n/A} (x, Q^2) = d^{p/A} (x, Q^2), & \ d^{n/A} (x, Q^2) = u^{p/A} (x, Q^2), \\ \bar{u}^{n/A} (x, Q^2) = \bar{d}^{p/A} (x, Q^2), & \ \bar{d}^{n/A} (x, Q^2) = \bar{u}^{p/A} (x, Q^2),
  \end{split}
\end{equation}
and $f_i^{n/A} = f_i^{p/A}$ for other flavours. This can be seen as an effective prescription, where the bound-nucleon PDFs should be understood as carrying information on the parton content of the ``average'' nucleon, used only to simplify the treatment of isospin dependence. In the region where $x \leq 1$, one can further write
\begin{equation}
  f_i^{p/A} (x, Q^2) = R_i^{p/A} (x, Q^2) f_i^{p} (x, Q^2),
  \label{eq:nucl_mod}
\end{equation}
where the nuclear modification factors $R_i^{p/A}$, given the free-proton PDFs $f_i^{p}$, now encode all the information on the partonic structure of nuclei.

It should be noted that the above steps can be taken without any loss of generality. In the end, if \emph{all} correlations between the proton and nuclear PDFs are correctly taken into account, it should not matter whether one parametrizes the absolute nuclear PDFs or the nuclear modifications.  One is simply mapping a set of unknown functions $f_i^A$ to the same number of functions $R_i^{p/A}$. In practice, however, simplifying assumptions are used in the nuclear-PDF analyses. We will use here the nuclear modifications from the EPPS16 analysis~\cite{Eskola:2016oht}, and for full consistency, we use the CT14 NLO free-proton PDFs~\cite{Dulat:2015mca} but also validate the robustness of the results by comparing to the CT18 NLO PDFs~\cite{Hou:2019efy}. This assumes that $R_i^{p/A}$ depend only on the nuclear mass number $A = Z + N$ (i.e.\ that there is no non-trivial isospin dependence in them) and that they are uncorrelated to $f_i^{p}$.\footnote{Note that $f_i^{p/A}$ and $f_i^A$ are still correlated to $f_i^{p}$ through Eqs.~\eqref{eq:nucl_mod} and~\eqref{eq:isospin}.} As mentioned in Section~\ref{sec:intro}, the latter is true as a first approximation due to the use of appropriate ratio observables in EPPS16, but we will discuss later in this article the validity of this assumption in the presence of increasingly precise electroweak data.

\begin{figure*}[t]
  \centering
  \includegraphics[width=1.66\columnwidth]{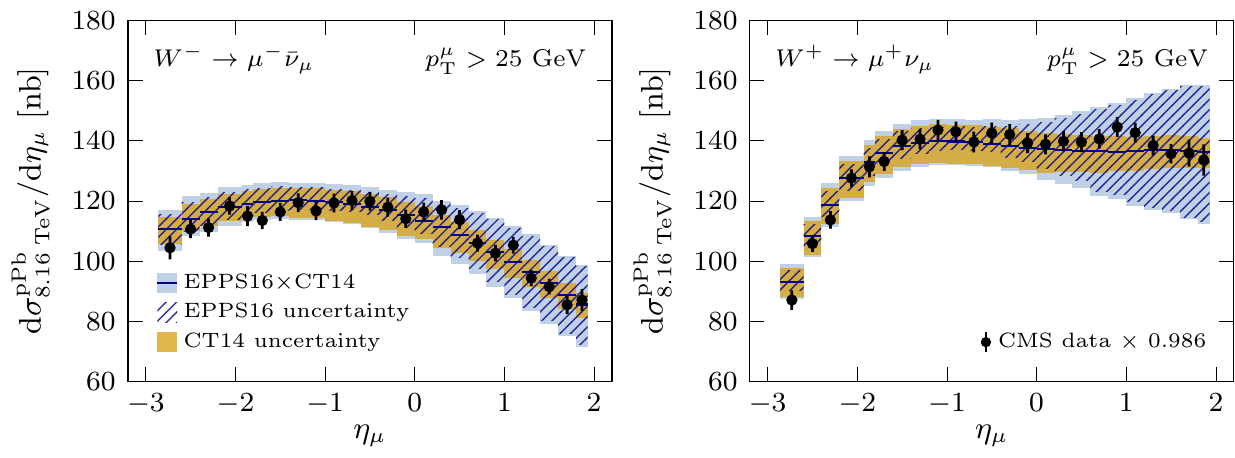}
  \caption{Lepton-rapidity differential $W^\pm$ production cross sections in p+Pb collisions at 8.16 TeV with a breakdown of the theoretical uncertainties (EPPS16$\times$CT14, light-blue boxes) into those from free-proton PDFs (CT14 NLO, yellow boxes) and from the nuclear modifications (EPPS16, blue hatching). The data from the CMS measurement~\cite{CMS:2019leu} are presented with black markers, scaled with the optimal normalization factor explained in the text.}
  \label{fig:CMS_Ws}
\end{figure*}

The advantage of this framework is that we can study the relative importance of the nuclear-modification and free-proton-PDF uncertainties in any observable of interest. Here, we study these in the context of $W^\pm$ production in p+Pb collisions at 8.16 TeV, as measured by the CMS Collaboration in the muon decay channel~\cite{CMS:2019leu}. The lepton-rapidity differential cross sections, with a cut on lepton transverse momentum $p_{\rm T}^\mu > 25\ {\rm GeV}$, is presented in Fig.~\ref{fig:CMS_Ws}. The theoretical next-to-leading order (NLO) perturbative QCD predictions are obtained with MCFM~\cite{Campbell:2015qma}, and the PDF uncertainties from EPPS16 and CT14 are calculated with the conventional asymmetric prescription at the 90\% confidence level. As can be seen from the figure, the baseline CT14 free-proton PDF errors (shown as yellow boxes) contribute significantly to the total theoretical uncertainty budget (light blue boxes) and can even exceed those from the EPPS16 nuclear modifications (blue hatching) in some bins. The smallness of nuclear-modification uncertainties in the negative (backward) rapidities originates from the good neutral and charged-current DIS constraints at the probed values of $x$. Going to positive (forward) rapidities, we enter the less-constrained small-$x$ region and the nuclear-modification uncertainties begin to grow.

As was shown already in Ref.~\cite{CMS:2019leu}, the agreement between the CMS measurement and the NLO predictions from EPPS16$\times$CT14 is excellent. The goodness of fit for this data set is given by
\begin{multline}
  \chi^2_C = \\(D - f_{\rm norm.} T)^{\rm T} C^{-1} (D - f_{\rm norm.} T) + \left( \frac{f_{\rm norm.} - 1}{\sigma_{\rm norm.}} \right)^2,
  \label{eq:chi2}
\end{multline}
where $D$ and $T$ are vectors of dimension $N_\text{data}$ containing the data and theory values, and we have extracted the normalization uncertainty $\sigma_{\rm norm.} = 3.5\%$ from the data covariance matrix $C$, thus avoiding the D'Agostini bias~\cite{DAgostini:1993arp}. By doing so, the optimal normalization factor $f_{\rm norm.}$ can be solved analytically, and in the figures we multiply the data with a factor
\begin{equation}
  1/f_{\rm norm.} = \frac{1 + \sigma_{\rm norm.}^2 T^{\rm T} C^{-1} T}{1 + \sigma_{\rm norm.}^2 D^{\rm T} C^{-1} T}.
  \label{eq:norm}
\end{equation}
Taking $T$ as the central prediction from EPPS16$\times$CT14, we then have $1/f_{\rm norm.} = 0.986$ and $\chi^2_{C} / N_\text{data} = 1.12$, confirming the visibly good data-to-theory agreement. When fitting to these data, we therefore do not expect the central nuclear PDFs to change much from the EPPS16 results, but as the experimental uncertainties are much smaller than the nuclear-modification uncertainties especially at forward rapidities, we can expect a significant reduction in the latter. The large baseline free-proton uncertainties can however affect the obtainable constraints and we need to find a way to either quantify or mitigate the impact.

\section{Theoretical covariance matrix}
\label{sec:thcov}

One way to quantify the impact of a certain theoretical source of uncertainty is to use the method of theoretical covariance matrix~\cite{NNPDF:2019ubu}. For the free-proton uncertainties, taken from the CT14 PDFs, this matrix is given by
\begin{multline}
  S^\text{CT14}_{ij} = \\\sum_k \frac{T_i[S_{\text{CT14}}^{k,+}] - T_i[S_{\text{CT14}}^{k,-}]}{2 \times 1.645} \frac{T_j[S_{\text{CT14}}^{k,+}] - T_j[S_{\text{CT14}}^{k,-}]}{2 \times 1.645},
\end{multline}
where the sum goes over the CT14 parameter eigendirections $k$ and $T_i[S_{\text{CT14}}^{k,\pm}]$ are the corresponding predictions for the $i$th data point with positive and negative parameter variations in that eigendirection (all calculated with the central EPPS16 nuclear modifications). The factors 1.645 in the denominators are used to scale the nominally 90\% confidence-level uncertainties of CT14 to a 68\% (one standard deviation) level in order not to overestimate their impact with respect to the experimental uncertainties.

\begin{figure}[b]
  \includegraphics[width=\columnwidth]{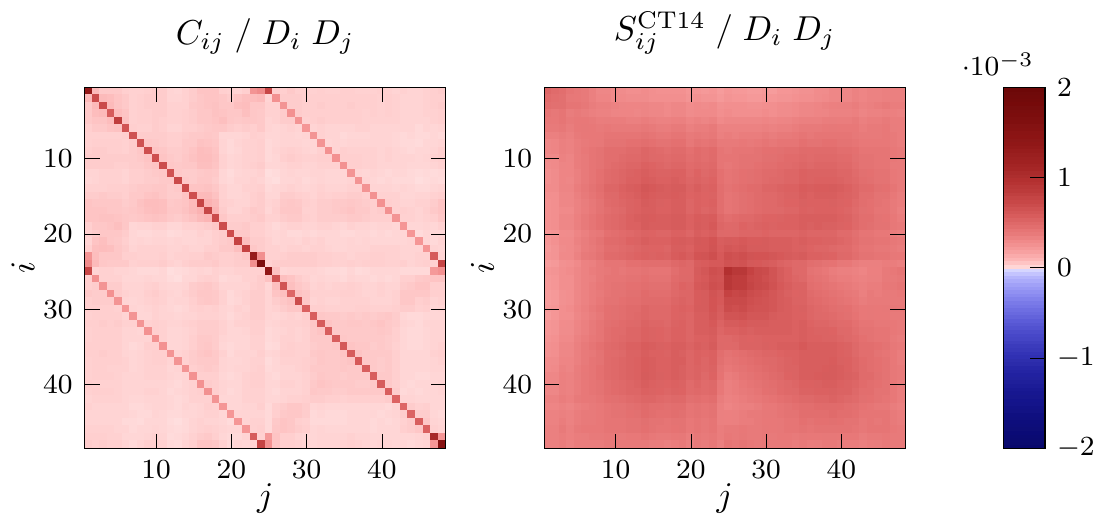}
  \caption{The experimental (excluding overall normalization uncertainty) and theoretical free-proton-PDF covariance matrices for the p+Pb $W^\pm$ measurement at 8.16 TeV. Indices $i,j$ follow the same ordering as the data points in Fig.~\ref{fig:CMS_Ws}, with the indices 1 through 24 corresponding to the $W^-$ production and 25 through 48 to $W^+$.}
  \label{fig:cov_mat}
\end{figure}

The CT14 theoretical covariance matrix is presented in Fig.~\ref{fig:cov_mat} with a comparison to the experimental covariance matrix. To ease the interpretation, we have excluded the large fully correlated luminosity component from the experimental covariance matrix, as in Eq.~\eqref{eq:chi2}, and divided each matrix element with the product of the corresponding data values. We see that the proton-PDF uncertainties are comparable or larger than the correlated non-luminosity experimental uncertainties but still mostly smaller than the combined statistical and non-luminosity systematical uncertainties in the diagonal elements. Clearly, the free-proton PDFs contribute a non-negligible uncertainty component to the nuclear-modification fitting.

With the theoretical CT14 proton-PDF uncertainties taken into account, the figure of merit for the nuclear-modification d.o.f.s takes the form~\cite{NNPDF:2019ubu}
\begin{multline}
  \chi^2_{C+S^\text{CT14}} = (D - f_{\rm norm.} T)^{\rm T} (C + S^\text{CT14})^{-1} (D - f_{\rm norm.} T) \\+ \left( \frac{f_{\rm norm.} - 1}{\sigma_{\rm norm.}} \right)^2,
  \label{eq:chi2thcov}
\end{multline}
and we find $\chi^2_{C+S^\text{CT14}} / N_\text{data} = 0.85$ for EPPS16. Comparing this to the value $\chi^2_{C} / N_\text{data} = 1.12$ for EPPS16$\times$ CT14, we see that the chosen proton PDFs can indeed have a significant impact on the level of agreement with the data.

Interestingly, the proton-PDF uncertainties are strongly positively correlated, behaving almost like an additional normalization uncertainty. It is exactly this positive correlation (and the positive correlation with the corresponding proton--proton cross section) which makes the uncertainty reduction with the ratios discussed in Section~\ref{sec:unc_red} possible. We note also that the optimal data normalization that we find for EPPS16$\times$CT14 from Eq.~\eqref{eq:norm} is $1/f_{\rm norm.} = 0.986$, well within the $3.5\%$ normalization uncertainty. This should be compared to the value of $0.960$ in the nCTEQ15WZ fit for these data~\cite{Kusina:2020lyz}. Since the proton PDFs contribute significantly to the normalization of the predictions, we can speculate whether the larger than $1\times\sigma_{\rm norm.}$ normalization shift in the nCTEQ15WZ analysis originates from the used free-proton baseline. This possibility is also corroborated by the fact that when testing the robustness of the results presented here by changing the free-proton PDFs to CT18 NLO, the main effect was a change in the normalization, with the optimal data-scaling factor $1/f_{\rm norm.}$ changing to a value $0.997$. The CT18 uncertainties were also observed to be slightly less correlated across different rapidities, but the uncertainties were found to be almost the same, and this had no impact on our conclusions.

\section{Reducing proton-PDF uncertainties}
\label{sec:unc_red}

As we have shown that the free-proton uncertainties are important in describing the p+Pb $W^\pm$ data, it makes sense to explore ways to reduce these uncertainties. Since the covariance matrix of the CMS measurement is available to us, we can propagate the data uncertainties to any desired observable keeping also track of the correlations by using
\begin{equation}
  C^\text{new} = J\,C\,J^{\rm T}, \label{eq:errorprop}
\end{equation}
where $J$ is the Jacobian of the transformation. We note that also perturbative higher-order corrections can (partially) cancel in many of the considered ratios, which supports their use in nuclear-PDF analyses, but the importance of missing higher orders is left outside the scope of this article.

\subsection{Self-normalized cross sections}
\label{sec:self-normalized}

\begin{figure*}[tb]
  \sidecaption
  \includegraphics[width=1.66\columnwidth]{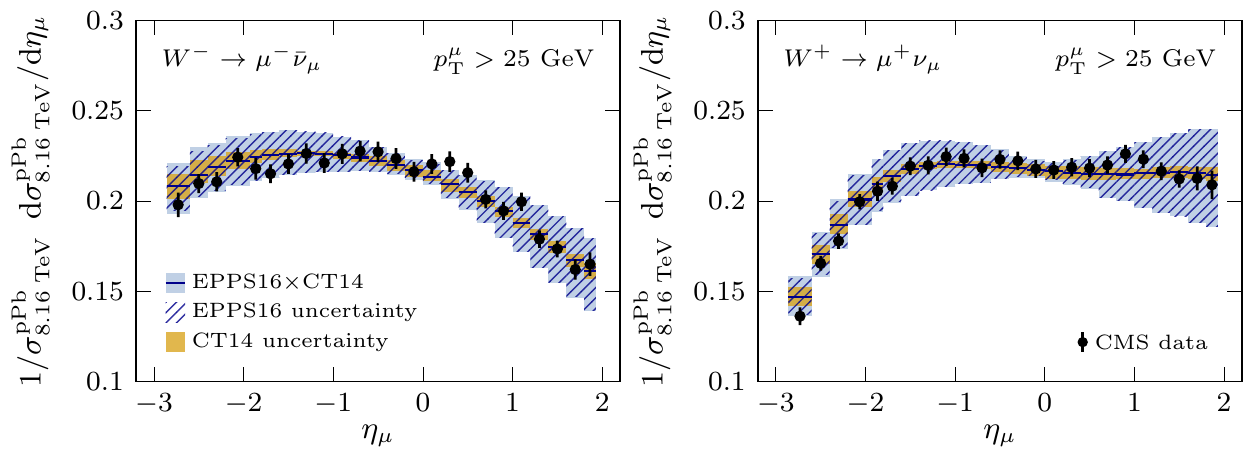}
  \caption{As Fig.~\ref{fig:CMS_Ws}, but now for the self-normal\-ized cross sections.}
  \label{fig:CMS_Ws_normalized}
\end{figure*}

\begin{figure}[htb]
  \includegraphics[width=\columnwidth]{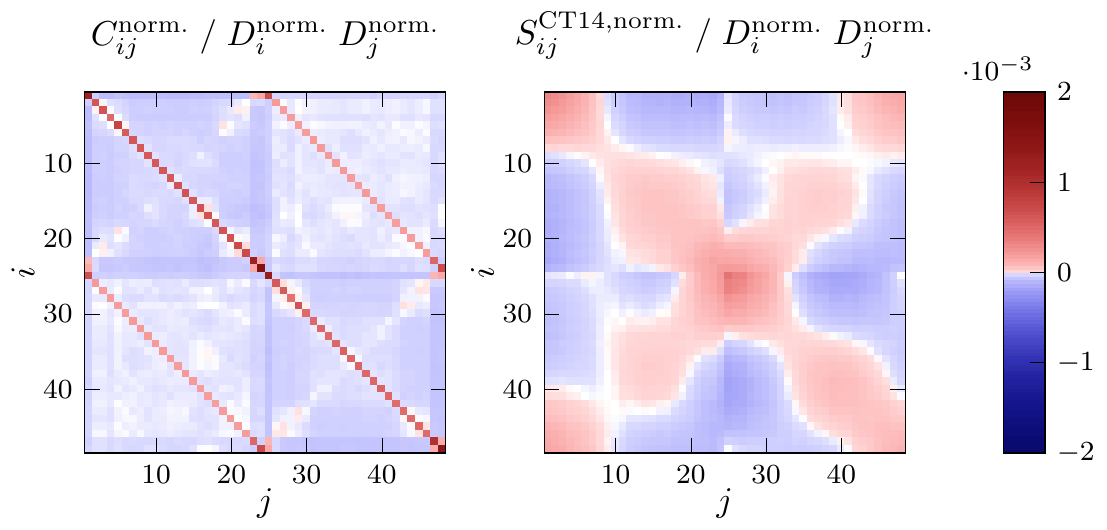}
  \caption{As Fig.~\ref{fig:cov_mat}, but now for the self-normalized cross sections. Indices $i,j$ follow the same ordering as the data points in Fig.~\ref{fig:CMS_Ws_normalized}, with the indices 1 through 24 corresponding to the $W^-$ production and 25 through 48 to $W^+$.}
  \label{fig:cov_mat_normalized}
\end{figure}

Since the free-proton PDF uncertainties were found to be strongly correlated, almost normalization-like, a viable option to reduce them is by self-normalizing the cross sections
\begin{equation}
  {\rm d}\sigma^{W^\pm,\text{norm.}}_{\rm pPb} / {\rm d}\eta_\mu = \frac{1}{\sigma^{W^\pm}_{\rm pPb}}\;{\rm d}\sigma^{W^\pm}_{\rm pPb} / {\rm d}\eta_\mu,
\end{equation}
where
\begin{equation}
  \sigma^{W^\pm}_{\rm pPb} = \int_{-2.86}^{1.93} {\rm d}\eta_\mu\;{\rm d}\sigma^{W^\pm}_{\rm pPb} / {\rm d}\eta_\mu
\end{equation}
are the fiducial integrated cross sections of each $W^\pm$ charge. We perform here the normalization for each charge separately, but it would be also possible to do this by dividing with the charge-summed integrated cross section. The obtained normalized cross sections are presented in Fig.~\ref{fig:CMS_Ws_normalized} and the experimental $C^{\rm norm.}$ (from Eq.~\eqref{eq:errorprop}) and theoretical $S^{\rm CT14,norm.}$ (calculated directly from the CT14 error sets) covariance matrices in Fig.~\ref{fig:cov_mat_normalized}. We observe a very good, but not perfect cancellation of the free-proton uncertainties, with the remaining CT14 uncertainties being largest at the large negative rapidities, $\eta_\mu < -1.93$.

\begin{figure*}[t]
  \sidecaption
  \includegraphics[width=1.66\columnwidth]{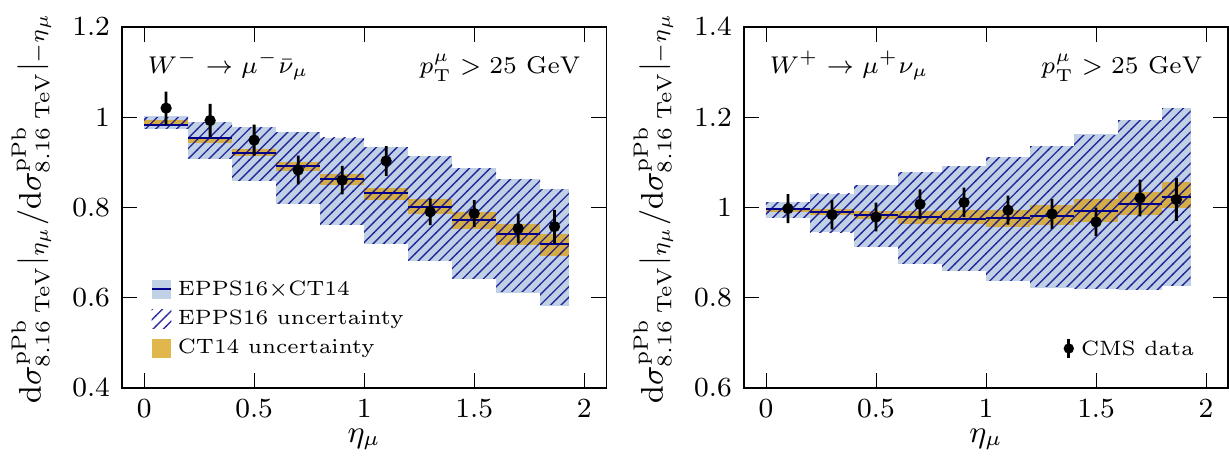}
  \caption{As Fig.~\ref{fig:CMS_Ws}, but now for the forward-to-backward ratios.}
  \label{fig:CMS_Ws_FB}
\end{figure*}

In addition to cancelling the normalization uncertainty, the self-normalization changes the correlation pattern of the remaining statistical and systematical experimental uncertainties, which become mostly anticorrelated across different rapidity bins. Importantly, even the originally uncorrelated (statistical) uncertainties become correlated in the self-normalized cross sections. Another important thing to notice here is that neither the experimental nor the theoretical covariance matrix is invertible, with $\det C^{\rm norm.} = \det S^{\rm CT14,norm.} = 0$. This simply follows from the fact that a self-normalized set of data forms an overdetermined system: given all but one data point, the last one can be solved from the requirement that the data integrate to one. Another way to see this is to notice that the self-normalization is not a bijection, with an immediate consequence that $\det J = 0$. For this property, one should fit to the self-normalized data by leaving one point out. Due to the fully correlated nature of the normalized data, it does not matter which data point is left out, manifesting the loss of information in the normalization.\footnote{We note, however, that e.g.\ in the case of the CMS measurement of self-normalized dijet cross sections in p+p and p+Pb collisions at 5.02~TeV~\cite{Sirunyan:2018qel}, which we have studied in Ref.~\cite{Eskola:2019dui}, the data correlations were not published and it is less clear how to treat the data statistically accurately in a fit. Without knowing the correlations, it \emph{would} matter which data point was left out.}

\subsection{Forward-to-backward ratios}

The forward-to-backward ratios
\begin{equation}
  R^{W^\pm}_{\rm FB} = \frac{{\rm d}\sigma^{W^\pm}_{\rm pPb} / {\rm d}\eta_\mu |_{\eta_\mu}}{{\rm d}\sigma^{W^\pm}_{\rm pPb} / {\rm d}\eta_\mu |_{-\eta_\mu}}
\end{equation}
have been considered earlier in Ref.~\cite{Paukkunen:2010qg}, where it was realised that they do not yield as good a cancellation of the free-proton uncertainties as e.g.\ the same ratios for $Z$-boson or dijet~\cite{Eskola:2013aya} production. Since the experimental acceptance is not symmetric with respect to the p+Pb center-of-mass frame, one has to drop part of the data points, in this case those for $\eta_\mu < -1.93$. Furthermore, by taking the ratio, the number of data points is still halved, leading to a significant loss of information. The remaining ten data points for each charge are shown in Fig.~\ref{fig:CMS_Ws_FB}, where we see that the proton-PDF cancellation is good, but starts to worsen towards larger rapidities.

\begin{figure}[htb]
  \includegraphics[width=\columnwidth]{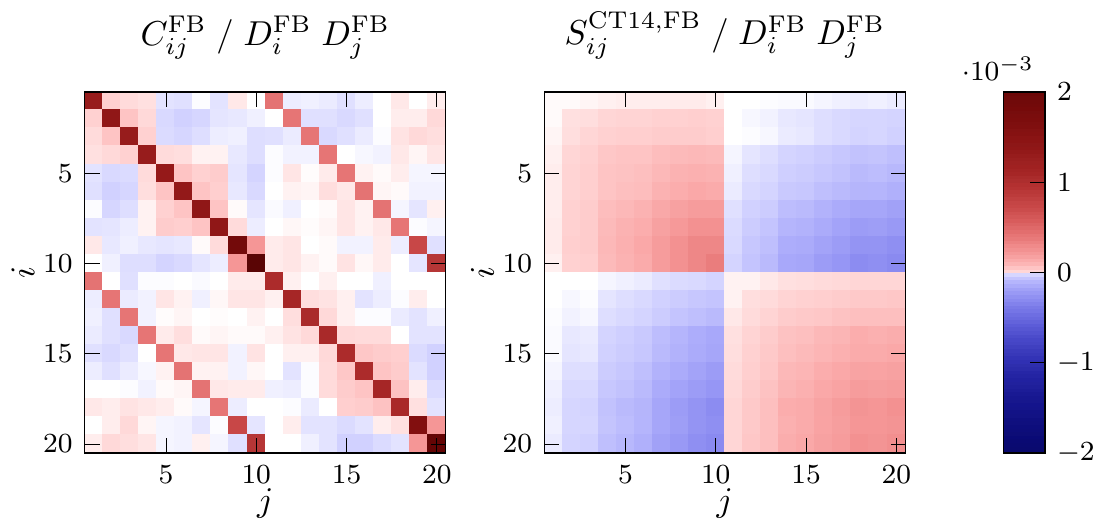}
  \caption{As Fig.~\ref{fig:cov_mat}, but now for the forward-to-backward ratio. Indices $i,j$ follow the same ordering as the data points in Fig.~\ref{fig:CMS_Ws_FB}, with the indices 1 through 10 corresponding to the $W^-$ production and 11 through 20 to $W^+$.}
  \label{fig:cov_mat_FB}
\end{figure}

\begin{figure*}[t]
  \centering
  \includegraphics[width=1.66\columnwidth]{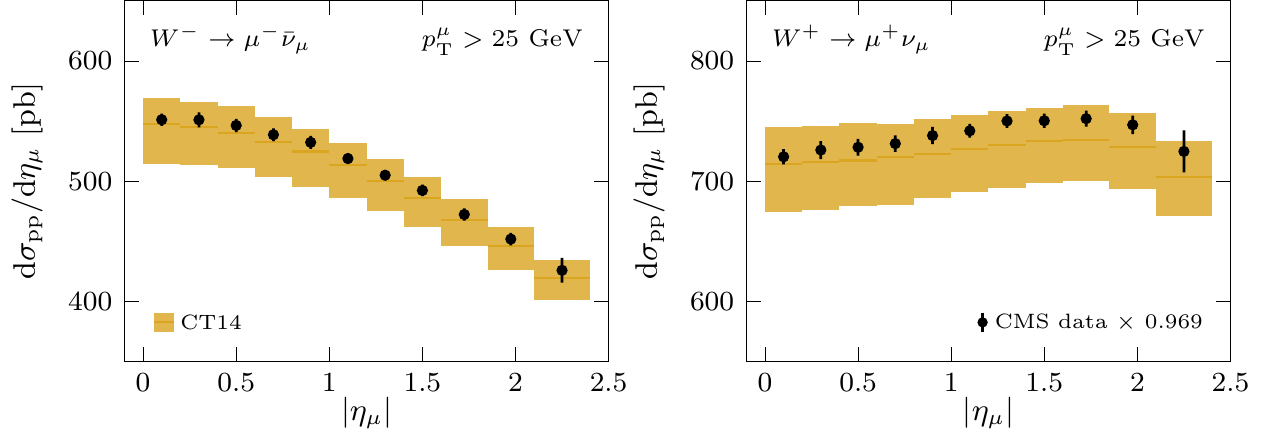}
  \caption{Lepton-rapidity differential $W^\pm$ production cross sections in p+p collisions at 8.0 TeV with theoretical uncertainties from the free-proton PDFs (CT14 NLO, yellow boxes). The data from the CMS Collaboration measurement~\cite{CMS:2016qqr} are presented with black markers, scaled with the optimal normalization factor.}
  \label{fig:CMS_Ws_pp}
\end{figure*}

This can be understood by taking the large-rapidity limit $\eta_\mu \gg 0$, where we can take the large momentum-fraction $x_1$ to probe only valence quarks and the small momentum-fraction $x_2$ then probes the sea quarks. At this limit, neglecting the Cabibbo suppressed quark-mixing effects and denoting $x_{1,2} \coloneqq x_{1,2}|_{\eta_\mu} = x_{2,1}|_{-\eta_\mu}$, we can approximate at leading order
\begin{equation}
  R^{W^-}_{\rm FB} \underset{\substack{x_1\,\text{large}\\x_2\,\text{small}}}{\overset{\eta_\mu \,\gg\, 0}{\approx}} \frac{Z R_{\bar{u}}^{p/A}(x_2) + N \frac{\bar{d}^p(x_2)}{\bar{u}^p(x_2)} R_{\bar{d}}^{p/A}(x_2)}{Z R_{d_{\rm V}}^{p/A}(x_1) + N \frac{u_{\rm V}^p(x_1)}{d_{\rm V}^p(x_1)} R_{u_{\rm V}}^{p/A}(x_1)} \label{eq:appr_FB_Wminus}
\end{equation}
and
\begin{equation}
  R^{W^+}_{\rm FB} \underset{\substack{x_1\,\text{large}\\x_2\,\text{small}}}{\overset{\eta_\mu \,\gg\, 0}{\approx}} \frac{Z R_{\bar{d}}^{p/A}(x_2) + N \frac{\bar{u}^p(x_2)}{\bar{d}^p(x_2)} R_{\bar{u}}^{p/A}(x_2)}{Z R_{u_{\rm V}}^{p/A}(x_1) + N \frac{d_{\rm V}^p(x_1)}{u_{\rm V}^p(x_1)} R_{d_{\rm V}}^{p/A}(x_1)}, \label{eq:appr_FB_Wplus}
\end{equation}
where we have suppressed for simplicity the relevant phase-space integrations and the scale-dependence of the PDFs. We note that this approximation is not exact in the data region as there can still be sizeable (but subleading) contributions also from the $\bar{c} + s$ and $c + \bar{s}$ channels~\cite{Kusina:2012vh}. In any case, we see that the forward-to-backward ratios depend on the free-proton PDFs through $u_{\rm V}/d_{\rm V}$ and $\bar{u}/\bar{d}$ ratios, which determine the relative size of the contributions from the different nuclear modifications and give a non-cancelling contribution to the theoretical uncertainty.

The resulting covariance matrices for the forward-to-backward ratios are presented in Fig.~\ref{fig:cov_mat_FB}, where we observe a positive correlation of the proton-PDF uncertainties between same-charge bins, but an anticorrelation between different charges. Indeed, one could reduce the proton-PDF uncertainties further by taking the forward-to-backward ratio of the differential cross section summed over the two charges, as considered in Ref.~\cite{CMS:2019leu}, but this leads to a further loss of information compared to taking the ratio separately for different charges, and the constraints for nuclear-PDF analyses are rather limited.

\subsection{Nuclear-modification ratios}

We now study the possibility of using nuclear-modification ratios to cancel free-proton uncertainties. For the 8.16 TeV p+Pb data, no same-energy p+p reference is available, but one could construct ``mixed-energy'' ratios
\begin{equation}
  R^{W^\pm}_{\rm pPb} = \frac{{\rm d}\sigma^{W^\pm}_{\rm pPb,\,8.16\,{\rm TeV}} / {\rm d}\eta_\mu}{A\,{\rm d}\sigma^{W^\pm}_{\rm pp,\,8.0\,{\rm TeV}} / {\rm d}\eta_\mu}
\end{equation}
with the p+p measurements at 8.0 TeV, where the probed $x$-ranges are almost the same between the two energies. We use here the measurements from the CMS Collaboration~\cite{CMS:2016qqr}, shown in Fig.~\ref{fig:CMS_Ws_pp} along with the predictions from the CT14 PDFs. Again, the $2.6\%$ normalization uncertainty is taken into account in presenting the data. The optimal shift, 0.969, is slightly larger than what we found for p+Pb. The agreement in normalization could be again improved by using CT18 PDFs, to 0.996, but the size of the PDF uncertainties stays almost the same.

The rapidity binning is the same in p+p and p+Pb measurements up to $|\eta_\mu| < 1.6$. For larger rapidities, we associate the p+Pb bins with the most-overlapping one in p+p. Therefore, for the p+Pb bins with $1.6 < |\eta_\mu| < 1.8$ we take the ratio with p+p bin $1.6 < |\eta_\mu| < 1.85$ and the p+p bin $1.85 < |\eta_\mu| < 2.1$ is used in obtaining three of the $R_{\rm pPb}$ bins: $1.8 < \eta_\mu < 1.93$, $-1.93 < \eta_\mu < -1.8$ and $-2.2 < \eta_\mu < -1.93$. Finally, for the p+Pb bin $-2.4 < \eta_\mu < -2.2$, the p+p bin $2.1 < |\eta_\mu| < 2.4$ is used. For $\eta_\mu < -2.4$ we run out of p+p bins and we discard the remaining two p+Pb data points for both $W^\pm$ charges. The loss of information is therefore slightly larger than in the self-normalized cross sections, but significantly smaller than in the forward-to-backward ratios.

\begin{figure}[htb]
  \includegraphics[width=\columnwidth]{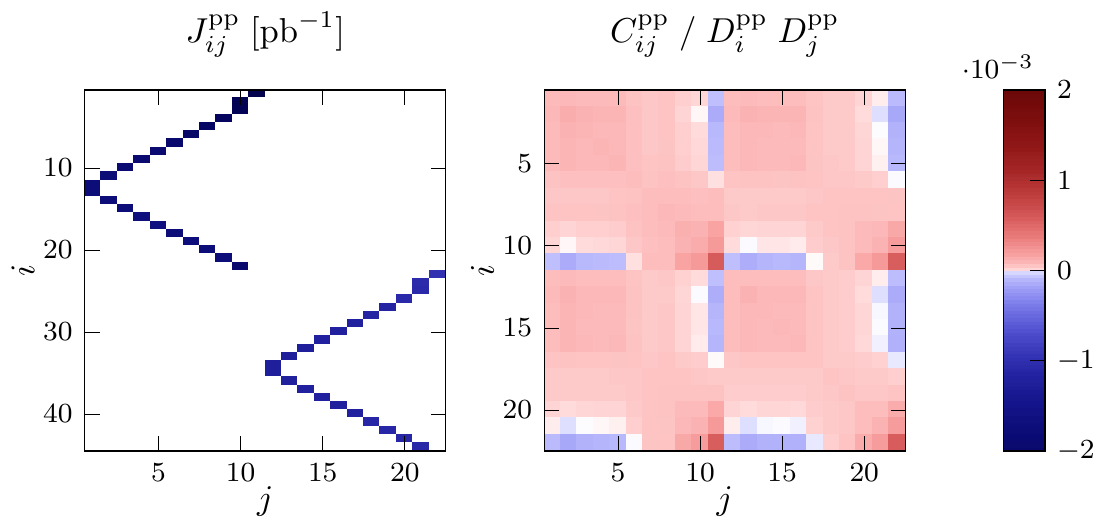}
  \caption{The Jacobian matrix $J^{\rm pp}_{ij}$ for propagating the p+p uncertainties into $R_{\rm pPb}$ uncertainties (cf.~Eq.~\eqref{eq:cov_mat_RpPb}) and the experimental (excluding overall normalization uncertainty) covariance matrix $C^{\rm pp}_{ij}$ for the p+p $W^\pm$ measurement at 8.0 TeV. Indices $i,j$ in $C^{\rm pp}_{ij}$ and the index $j$ in $J^{\rm pp}_{ij}$ follow the ordering of the data points in Fig.~\ref{fig:CMS_Ws_pp}, with the indices 1 through 11 corresponding to the $W^-$ production and 12 through 22 to $W^+$, and the index $i$ in $J^{\rm pp}_{ij}$ follows the ordering in Fig.~\ref{fig:CMS_Ws_RpPb}, with the indices 1 through 22 corresponding to the $W^-$ production and 23 through 44 to $W^+$.}
  \label{fig:cov_mat_pp}
\end{figure}

\begin{figure*}[t]
  \sidecaption
  \includegraphics[width=1.66\columnwidth]{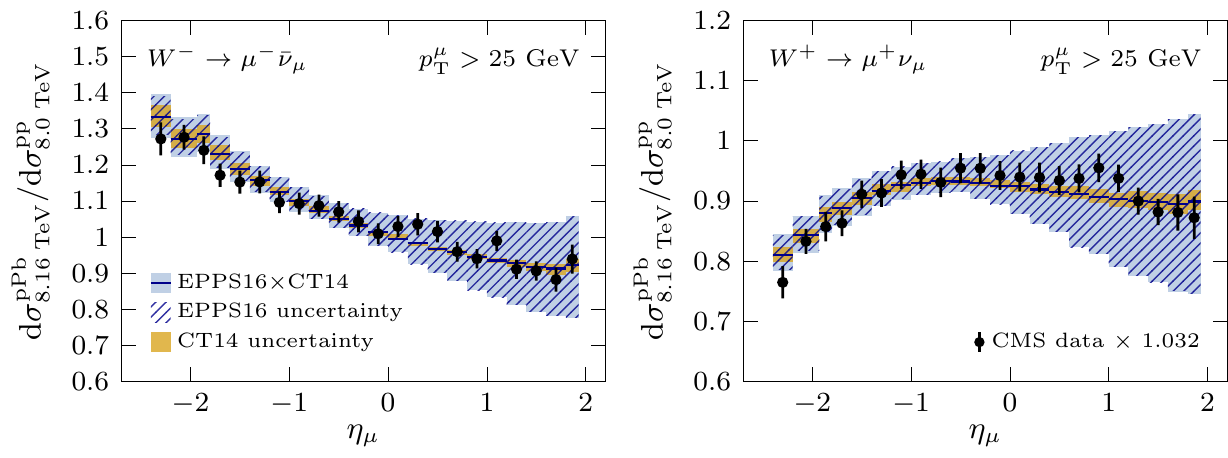}
  \caption{As Fig.~\ref{fig:CMS_Ws}, but now for the nuclear-modification ratio with the p+p reference taken from Ref.~\cite{CMS:2016qqr}.}
  \label{fig:CMS_Ws_RpPb}
\end{figure*}

Since the correlations between the p+Pb and p+p measurements are not known, the covariance matrix for the ratio is calculated with
\begin{equation}
  C^{R_{\rm pPb}} = J^{\rm pPb}\,C^{\rm pPb}\,(J^{\rm pPb})^{\rm T} + J^{\rm pp}\,C^{\rm pp}\,(J^{\rm pp})^{\rm T}.
  \label{eq:cov_mat_RpPb}
\end{equation}
This is a conservative estimate: in a direct experimental analysis some of the systematic uncertainties could be cancelled in the ratio. The Jacobian $J^{\rm pp}$ and the covariance matrix $C^{\rm pp}$ for the p+p data are presented in Fig.~\ref{fig:cov_mat_pp}, visualising also how each p+p point contributes to multiple $R_{\rm pPb}$ bins. The correlations arising from this are then taken correctly into account in Eq.~\eqref{eq:cov_mat_RpPb}.

\begin{figure}[htb]
  \includegraphics[width=\columnwidth]{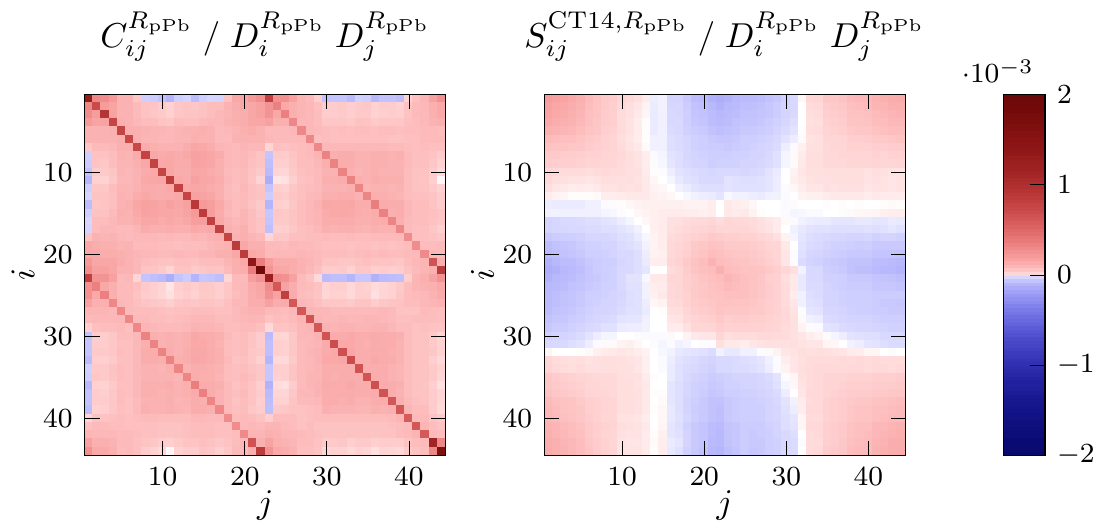}
  \caption{As Fig.~\ref{fig:cov_mat}, but now for the nuclear-modification ratio with the p+p reference taken from Ref.~\cite{CMS:2016qqr}. Indices $i,j$ follow the same ordering as the data points in Fig.~\ref{fig:CMS_Ws_RpPb}, with the indices 1 through 22 corresponding to the $W^-$ production and 23 through 44 to $W^+$.}
  \label{fig:cov_mat_RpPb}
\end{figure}

The obtained mixed-energy nuclear-modification ratios and the corresponding experimental and theoretical covariance matrices are presented in Figures~\ref{fig:CMS_Ws_RpPb} and~\ref{fig:cov_mat_RpPb}, respectively. The data are again well described by the EPPS16$\times$CT14 predictions, but due to the larger optimal downward normalization shift in p+p compared to p+Pb, we find the optimal shift for the nuclear-modification ratio to be $1/f_{\rm norm.} = 1.032$, still within the combined normalization uncertainty of 4.36\%. Compared to the previously discussed ratios, we can expect a more ``local'' cancellation of the proton-PDF dependence. However, since we use different collision energies for p+p and p+Pb and the rapidity binning does not exactly match outside mid-rapidity, the probed $x$ regions in the ratio can be slightly different, which can make the proton-PDF cancellation less than perfect.\footnote{In a direct experimental measurement of the ratio, one could consider binning the data in a shifted rapidity variable $y_{\rm ref}$ as in Ref.~\cite{Arleo:2015dba} to minimize the effect of using different energies.} We observe still a very good cancellation, comparable or better than in the previous ratios, and the CT14 uncertainties in Fig.~\ref{fig:cov_mat_RpPb} are now clearly smaller than the diagonal elements of the experimental covariance matrix in all bins (note that we have also here omitted the overall experimental normalization uncertainty from the presentation of the matrix, in accordance with Eq.~\eqref{eq:chi2}). Consequently, the free-proton PDFs have smaller impact on the agreement with data, as we find $\chi^2_{C} / N_\text{data} = 0.77$ with the pure experimental uncertainties and $\chi^2_{C+S^\text{CT14}} / N_\text{data} = 0.75$ after taking the CT14 uncertainties into account.

The cancellation somewhat deteriorates towards larger rapidities. In the far-backward region, the momentum-fraction from the nuclear side $x_2$ is large and we can approximate, at leading order and neglecting the small shifts in the momentum fractions due to the different energies,
\begin{equation}
  R^{W^-}_{\rm pPb} \underset{x_2\,\text{large}}{\overset{\eta_\mu \,\ll\, 0}{\approx}} \frac{Z}{A} R_{d_{\rm V}}^{p/{\rm Pb}}(x_2) + \frac{N}{A} \frac{u_{\rm V}^p(x_2)}{d_{\rm V}^p(x_2)} R_{u_{\rm V}}^{p/{\rm Pb}}(x_2)
  \label{eq:RpPbminusbwd}
\end{equation}
and
\begin{equation}
  R^{W^+}_{\rm pPb} \underset{x_2\,\text{large}}{\overset{\eta_\mu \,\ll\, 0}{\approx}} \frac{Z}{A} R_{u_{\rm V}}^{p/{\rm Pb}}(x_2) + \frac{N}{A} \frac{d_{\rm V}^p(x_2)}{u_{\rm V}^p(x_2)} R_{d_{\rm V}}^{p/{\rm Pb}}(x_2),
  \label{eq:RpPbplusbwd}
\end{equation}
where we see that the proton $u_{\rm V}/d_{\rm V}$ ratio again sets the limit to how well the proton-PDF uncertainties are cancelled. Note that in Eq.~\eqref{eq:RpPbminusbwd} we have the ratio $u_{\rm V}/d_{\rm V}$, leading to an enhancement at the probed backward rapidities, whereas in Eq.~\eqref{eq:RpPbplusbwd} we have its reciprocal, leading to a suppression, even in absence of nuclear modifications.

In the far-forward region the nuclear momentum-fraction $x_2$ is small and we have
\begin{equation}
  R^{W^-}_{\rm pPb} \underset{x_2\,\text{small}}{\overset{\eta_\mu \,\gg\, 0}{\approx}} \frac{Z}{A} R_{\bar{u}}^{p/{\rm Pb}}(x_2) + \frac{N}{A} \frac{\bar{d}^p(x_2)}{\bar{u}^p(x_2)} R_{\bar{d}}^{p/{\rm Pb}}(x_2)
  \label{eq:RpPbplusfwd}
\end{equation}
and
\begin{equation}
  R^{W^+}_{\rm pPb} \underset{x_2\,\text{small}}{\overset{\eta_\mu \,\gg\, 0}{\approx}} \frac{Z}{A} R_{\bar{d}}^{p/{\rm Pb}}(x_2) + \frac{N}{A} \frac{\bar{u}^p(x_2)}{\bar{d}^p(x_2)} R_{\bar{u}}^{p/{\rm Pb}}(x_2).
  \label{eq:RpPbminusfwd}
\end{equation}
At this limit, we see from Fig.~\ref{fig:CMS_Ws_RpPb} that the ratios of both charges approach the value 0.9, reflecting the fact that at these large scales, one is probing an almost flavour symmetric quark sea generated through $g \rightarrow q\bar{q}$ splittings, and therefore also the probed nuclear modifications are almost the same.

\subsection{Charge asymmetries}
\label{sec:asymmetries}

\begin{figure}[b]
  \centering
  \includegraphics[width=\columnwidth]{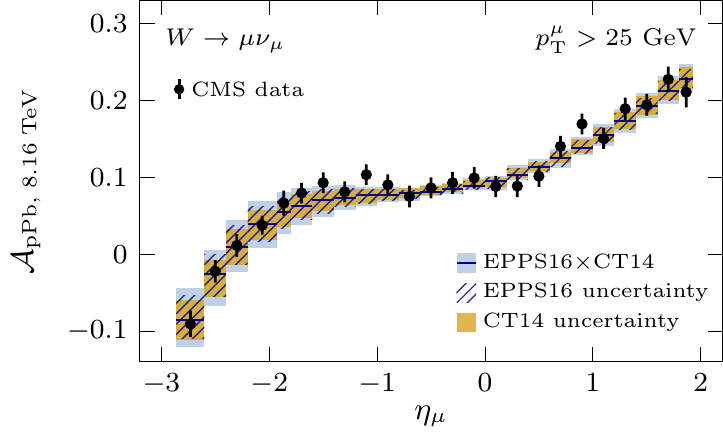}
  \caption{The $W^\pm$ production charge asymmetry, with a breakdown of theory uncertainties as in Fig.~\ref{fig:CMS_Ws}.}
  \label{fig:CMS_Ws_charge_asymmetry}
\end{figure}

As discussed already in Ref.~\cite{Paukkunen:2010qg}, the traditional charge asymmetry
\begin{equation}
  \mathcal{A}_{\rm pPb} = \frac{{\rm d}\sigma^{W^+}_{\rm pPb} / {\rm d}\eta_\mu - {\rm d}\sigma^{W^-}_{\rm pPb} / {\rm d}\eta_\mu}{{\rm d}\sigma^{W^+}_{\rm pPb} / {\rm d}\eta_\mu + {\rm d}\sigma^{W^-}_{\rm pPb} / {\rm d}\eta_\mu}
\end{equation}
is very sensitive to the free-proton uncertainties. As shown in Fig.~\ref{fig:CMS_Ws_charge_asymmetry}, the excellent data-to-theory agreement continues to be valid also in this observable, but as now the free-proton and nuclear-modification uncertainties are of the same order, one is probing a non-trivial combination of the two, and the usefulness for nuclear-PDF fits is rather limited. In particular, at large positive rapidities this observable probes mostly the $u_{\rm V} - d_{\rm V}$ asymmetry in proton~\cite{Kusina:2016fxy}, with the nuclear uncertainties having a strong cancellation.

\begin{figure*}[t]
  \sidecaption
  \includegraphics[width=0.81\columnwidth]{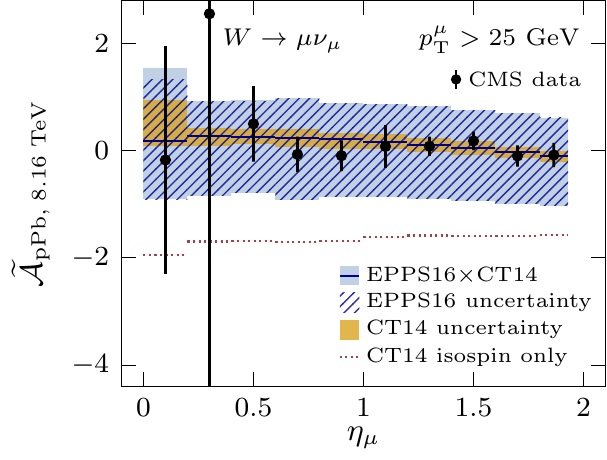}
  \includegraphics[width=0.81\columnwidth]{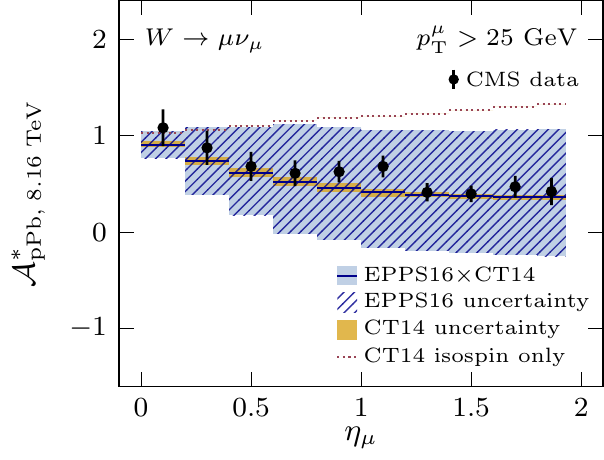}
  \caption{As Fig.~\ref{fig:CMS_Ws}, but now for the alternative charge-asymmetry ratios.}
  \label{fig:CMS_Ws_other_asymmetries}
\end{figure*}

It is, however, possible to construct asymmetries with more direct sensitivity to the nuclear modifications. In Ref.~\cite{Paukkunen:2010qg}, a charge ratio of forward--backward differences
\begin{equation}
  \tilde{\mathcal{A}}_{\rm pPb} = \frac{{\rm d}\sigma^{W^+}_{\rm pPb} / {\rm d}\eta_\mu |_{\eta_\mu} - {\rm d}\sigma^{W^+}_{\rm pPb} / {\rm d}\eta_\mu |_{-\eta_\mu}}{{\rm d}\sigma^{W^-}_{\rm pPb} / {\rm d}\eta_\mu |_{\eta_\mu} - {\rm d}\sigma^{W^-}_{\rm pPb} / {\rm d}\eta_\mu |_{-\eta_\mu}},
  \label{eq:another_asymmetry}
\end{equation}
shown in Fig.~\ref{fig:CMS_Ws_other_asymmetries} (left), was proposed, motivated by the finding that for cross sections differential in the $W^\pm$ boson rapidity, the proton-PDF uncertainties cancel extremely well in this quantity. Here, with the experimentally measurable lepton rapidity, we find the cancellation to be slightly worse, but it still gives far better access to the nuclear modifications than the traditional charge asymmetry. In particular, the measured data differ significantly from the predictions with free-proton PDFs taking into account the isospin effects only, i.e.\ neglecting the nuclear modifications in the bound-nucleon PDFs. Close to midrapidity this observable is experimentally problematic since the denominator approaches zero, and we find with the linear error propagation the statistics to be insufficient for any constraints at $|\eta_\mu| < 0.4$.

We can now use our knowledge of the proton-PDF correlations to our advantage. As can be seen from Figures~\ref{fig:cov_mat_normalized} and~\ref{fig:CMS_Ws_RpPb}, after taking away the overall normalization-like contribution, there is an anticorrelation in the proton-PDF uncertainties between the same-charge forward and backward cross sections. This anticorrelation is reflected also in the imperfect free-proton-PDF cancellation in the forward-to-backward ratios. Conversely, and quite unexpectedly, there appears to be a \emph{positive} correlation between the forward production of one charge and the backward production of the other and an anticorrelation between the two charges at the same rapidity. Based on the approximation in Eqs.~\eqref{eq:RpPbminusbwd} through~\eqref{eq:RpPbminusfwd}, this appears to be possible only if the ratio $\bar{u}^p(x_2)/\bar{d}^p(x_2)$ at small $x_2$ and $u_{\rm V}^p(x_1)/d_{\rm V}^p(x_1)$ at large $x_1$ are positively correlated. We find this to be true at the EW scale for CT14 (and also for CT18, MSHT20~\cite{Bailey:2020ooq} and NNPDF4.0~\cite{Ball:2021leu}) as long as $x_2 < 0.03$, independently of $x_1$.

With this information in mind, we construct here a new forward-to-backward ratio of rapidity-mirrored charge difference
\begin{equation}
  \mathcal{A}_{\rm pPb}^* = \frac{{\rm d}\sigma^{W^+}_{\rm pPb} / {\rm d}\eta_\mu |_{\eta_\mu} - {\rm d}\sigma^{W^-}_{\rm pPb} / {\rm d}\eta_\mu |_{-\eta_\mu}}{{\rm d}\sigma^{W^+}_{\rm pPb} / {\rm d}\eta_\mu |_{-\eta_\mu} - {\rm d}\sigma^{W^-}_{\rm pPb} / {\rm d}\eta_\mu |_{\eta_\mu}},
  \label{eq:newasym}
\end{equation}
shown in Fig.~\ref{fig:CMS_Ws_other_asymmetries}~(right). The free-proton-PDF uncertainties in this observable are negligible and it avoids the problem of vanishing denominator that appeared in Eq.~\eqref{eq:another_asymmetry} as the $W^+$ cross section is always larger than the $W^-$ one. Within the approximation used in Eqs.~\eqref{eq:appr_FB_Wminus} and~\eqref{eq:appr_FB_Wplus}, it can be written in the large-rapidity limit as
\begin{strip}
  \rule[-1ex]{\columnwidth}{0.4pt}\rule[-1ex]{0.4pt}{1.4ex}
  \begin{equation}
    \mathcal{A}_{\rm pPb}^* \underset{\substack{x_1\,\text{large}\\x_2\,\text{small}}}{\overset{\eta_\mu \,\gg\, 0}{\approx}} \frac{Z \big[ R_{\bar{d}}^{p/A}(x_2) - \frac{\bar{u}^p(x_2)}{\bar{d}^p(x_2)} \frac{d_{\rm V}^p(x_1)}{u_{\rm V}^p(x_1)} R_{d_{\rm V}}^{p/A}(x_1) \big] + N \frac{\bar{u}^p(x_2)}{\bar{d}^p(x_2)} \big[ R_{\bar{u}}^{p/A}(x_2) - R_{u_{\rm V}}^{p/A}(x_1) \big]}{Z \big[ R_{u_{\rm V}}^{p/A}(x_1) - \frac{\bar{u}^p(x_2)}{\bar{d}^p(x_2)} \frac{d_{\rm V}^p(x_1)}{u_{\rm V}^p(x_1)} R_{\bar{u}}^{p/A}(x_2) \big] + N \frac{d_{\rm V}^p(x_1)}{u_{\rm V}^p(x_1)} \big[ R_{d_{\rm V}}^{p/A}(x_1) - R_{\bar{d}}^{p/A}(x_2) \big]}.
  \end{equation}
  \hfill\rule[1ex]{0.4pt}{1.4ex}\rule[2.3ex]{\columnwidth}{0.4pt}
\end{strip}%
The proton-PDF cancellation is still not exact, but we see that unlike in the forward-to-backward ratios where the proton-PDF correlations lead to a strictly additive behaviour of the uncertainties, in this asymmetry the correlations contribute in a destructive way in the terms proportional to $Z$. There is also large cancellation of the $N$-proportional terms in the differences, and at the ``isospin only'' limit these terms vanish completely. Indeed, one notices that in a (hypothetical) free-proton--free-neutron collision at leading order the dominant $u+\bar{d}$ and $\bar{u}+d$ contributions cancel in the difference ${\rm d}\sigma^{W^+}_{\rm pn} / {\rm d}\eta_\mu |_{\eta_\mu} - {\rm d}\sigma^{W^-}_{\rm pn} / {\rm d}\eta_\mu |_{-\eta_\mu}$. As ${\rm d}\sigma^{W^+}_{\rm pp} / {\rm d}\eta_\mu |_{\eta_\mu} - {\rm d}\sigma^{W^-}_{\rm pp} / {\rm d}\eta_\mu |_{-\eta_\mu}$ is instead forward-to-backward symmetric, the asymmetry in the ``isospin only'' limit is then approximately one.

When the nuclear modifications are taken into account, this asymmetry is seen to probe a fairly non-trivial combination of different nuclear-modification terms, and the achievable constraints will depend on their relative uncertainties. Similarly to the forward-to-backward ratios, both a nuclear suppression in the sea quarks at small $x$ and an enhancement in the valence quarks at large $x_1$ cause a suppression in the ratio, explaining the strong deviation from the ``isospin only'' prediction.

For the other alternative asymmetry one finds that
\begin{strip}
  \rule[-1ex]{\columnwidth}{0.4pt}\rule[-1ex]{0.4pt}{1.4ex}
  \begin{equation}
    \tilde{\mathcal{A}}_{\rm pPb} \underset{\substack{x_1\,\text{large}\\x_2\,\text{small}}}{\overset{\eta_\mu \,\gg\, 0}{\approx}} \frac{Z \big[ R_{\bar{d}}^{p/A}(x_2) - R_{u_{\rm V}}^{p/A}(x_1) \big] + N \big[ \frac{\bar{u}^p(x_2)}{\bar{d}^p(x_2)} R_{\bar{u}}^{p/A}(x_2) - \frac{d_{\rm V}^p(x_1)}{u_{\rm V}^p(x_1)} R_{d_{\rm V}}^{p/A}(x_1) \big]}{Z \frac{\bar{u}^p(x_2)}{\bar{d}^p(x_2)} \frac{d_{\rm V}^p(x_1)}{u_{\rm V}^p(x_1)} \big[ R_{\bar{u}}^{p/A}(x_2) - R_{d_{\rm V}}^{p/A}(x_1) \big] + N \big[ \frac{d_{\rm V}^p(x_1)}{u_{\rm V}^p(x_1)} R_{\bar{d}}^{p/A}(x_2) - \frac{\bar{u}^p(x_2)}{\bar{d}^p(x_2)} R_{u_{\rm V}}^{p/A}(x_1) \big]},
  \end{equation}
  % \hfill\rule[1ex]{0.4pt}{1.4ex}\rule[2.3ex]{\columnwidth}{0.4pt}
\end{strip}%
where the proton-PDF uncertainty reduction is again apparent in the $Z$-proportional terms. It is now these terms that have the large cancellation and vanish at the ``isospin only'' limit, owing to the forward-to-backward symmetry of the p+p system, and the asymmetry reduces to minus one in this approximation.

\begin{figure}[htb]
  \includegraphics[width=\columnwidth]{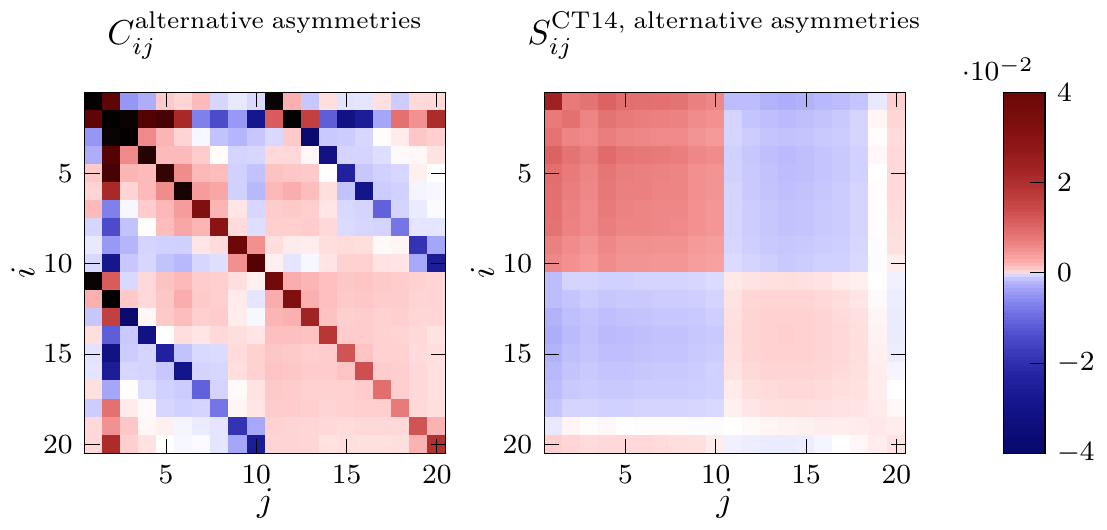}
  \caption{As Fig.~\ref{fig:cov_mat}, but now for the alternative asymmetries shown in Fig.~\ref{fig:CMS_Ws_other_asymmetries}. Note that here we do not normalize by the data values as they can be smaller than the uncertainties in $\tilde{\mathcal{A}}_{\rm pPb}$. The scale of the heat map is thus different from the other figures. Indices $i,j$ follow the same ordering as the data points in Fig.~\ref{fig:CMS_Ws_other_asymmetries}, with the indices 1 through 10 corresponding to $\tilde{\mathcal{A}}_{\rm pPb}$ and 11 through 20 to $\mathcal{A}_{\rm pPb}^*$.}
  \label{fig:cov_mat_other_asymmetries}
\end{figure}

Since the same cross sections are used in each of them, the experimental uncertainties of the different asymmetries are strongly correlated, as can be seen from Fig.~\ref{fig:cov_mat_other_asymmetries} for the alternative asymmetries, and must be taken into account when performing a simultaneous fit. We also note that in constructing these ratios, like in the case of forward-to-backward ratios, one again had to discard part of the p+Pb data, which might lead to reduced constraints.

\section{Hessian reweighting}
\label{sec:rw}

\begin{figure*}[tp]
  \centering
  \includegraphics[width=1.9\columnwidth]{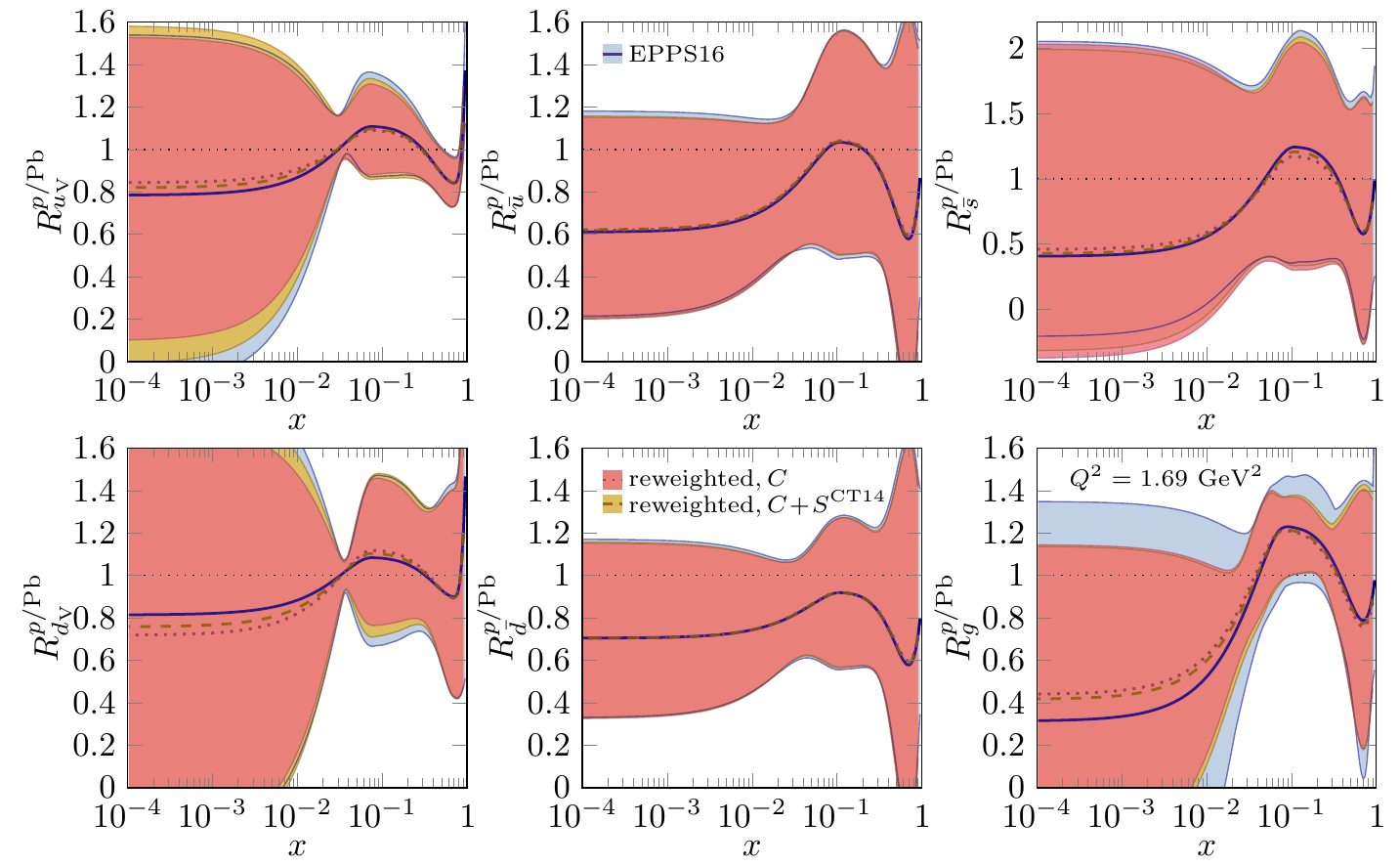}
  \caption{The results of reweighting the EPPS16 (blue bands and solid lines) nuclear modifications with the $W^\pm$ absolute cross sections, both without (``$C$'', red bands and dotted lines) and with (``$C+S^\text{CT14}$'', yellow bands and dashed lines) the CT14 uncertainties included.}
  \label{fig:rwwithabsxsec}
\end{figure*}

To further test the importance of the free-proton uncertainties in a nuclear-PDF fit, we have employed here the Hessian PDF-reweighting method~\cite{Paukkunen:2013grz,Paukkunen:2014zia,Eskola:2019dui,Schmidt:2018hvu,Hou:2019gfw}. This proceeds by supplementing Eqs.~\eqref{eq:chi2} and~\eqref{eq:chi2thcov} with a penalty term $P_\text{EPPS16}$, such that
\begin{equation}
  \Delta\chi^2_\text{total,\,$C$}(\vec{z}) = \chi^2_{C}(\vec{z}) + P_\text{EPPS16}(\vec{z})
  \label{eq:chi2rw}
\end{equation}
and
\begin{equation}
  \Delta\chi^2_\text{total,\,$C+S^\text{CT14}$}(\vec{z}) = \chi^2_{C+S^\text{CT14}}(\vec{z}) + P_\text{EPPS16}(\vec{z})
  \label{eq:chi2rwthcov}
\end{equation}
approximate the change in the total figure of merit of a global analysis, where the CMS data would be included on top of those in the EPPS16 analysis, as a function of the eigenparameters $\vec{z}$ of the EPPS16 analysis, either without or with the CT14 theoretical covariance matrix included. As in Ref.~\cite{Eskola:2019dui}, we take into account the leading non-quadratic terms by taking
\begin{equation}
  P_\text{EPPS16}(\vec{z}) = \sum_k (a_k z_k^2 + b_k z_k^3)
\end{equation}
and in $\chi^2_{C}$ and $\chi^2_{C+S^\text{CT14}}$ by including the first non-linear terms in
\begin{equation}
  T(\vec{z}) = T_0 + \sum_k (d_k z_k + e_k z_k^2),
\end{equation}
where the central prediction $T_0$ as well as the coefficients $a_k, b_k \in \mathbb{R}$ and $d_k, e_k \in \mathbb{R}^{N_\text{data}}$ are calculated with the information provided in the EPPS16 analysis~\cite{Eskola:2016oht}. By minimizing Eq.~\eqref{eq:chi2rw} or \eqref{eq:chi2rwthcov}, one finds the new, updated best estimate of the nuclear PDFs, and, by diagonalizing the Hessian matrix at the found minimum, one can define the new error sets, as explained in Ref.~\cite{Eskola:2019dui}.

Fig.~\ref{fig:rwwithabsxsec} shows the nuclear modifications after reweighting with the absolute cross sections from Fig.~\ref{fig:CMS_Ws}, presented at the EPPS16 parametrization scale $Q^2 = 1.69\ {\rm GeV}$, both without (``$C$'') and with (``$C+S^\text{CT14}$'') the CT14 uncertainties included. The results might appear surprising, in the sense that the $W^\pm$ production data which are directly sensitive to the $u_{\rm V}$ and $d_{\rm V}$ distributions of the nucleus do not put stronger constraints on the valence-quark nuclear modification factors. This is due to the existing constraints from CHORUS neutrino--nucleus DIS data~\cite{CHORUS:2005cpn} included in the EPPS16 analysis, on top of which these additional constraints from $W^\pm$ bosons are rather modest. For sea quarks, we find the constraints at the parametrization scale to be negligible. This originates from the EW-scale quark sea being dictated by the parametrization-scale gluons. For this reason, at the parametrization scale, we obtain the largest constraints for the gluon nuclear modifications, consistent with those~\cite{Eskola:2019bgf,Kusina:2017gkz,Eskola:2019dui} from dijet and D$^0$-meson production measurements~\cite{Sirunyan:2018qel,Aaij:2017gcy}. We note that these hadronic observables favour a larger small-$x$ suppression compared to EPPS16, whereas the opposite is true for the $W^\pm$ bosons. Everything is still consistent within uncertainties and the universality of nuclear PDFs appears to hold.

While the impact of the data on the valence modifications is not very large, there is still some dependence in the results on whether the proton-PDF uncertainties are taken into account or not. Based on the very large free-proton-PDF uncertainties we saw in Fig.~\ref{fig:CMS_Ws}, one could have expected these to have even larger impact in the reweighting. The smallness of the impact can be understood based on our observation that the dominant proton-PDF uncertainty in the absolute cross sections was normalization-like. Combined with the significant experimental luminosity uncertainty, it therefore has a minor impact on the nuclear modifications. However, while the difference in the ``$C$'' versus ``$C+S^\text{CT14}$'' might not appear large, it should be noticed that including the proton-PDF uncertainties washes away a large part of the valence-quark flavour-separation constraints that we would have otherwise had from these data.

\begin{figure*}[tp]
  \centering
  \includegraphics[width=1.9\columnwidth]{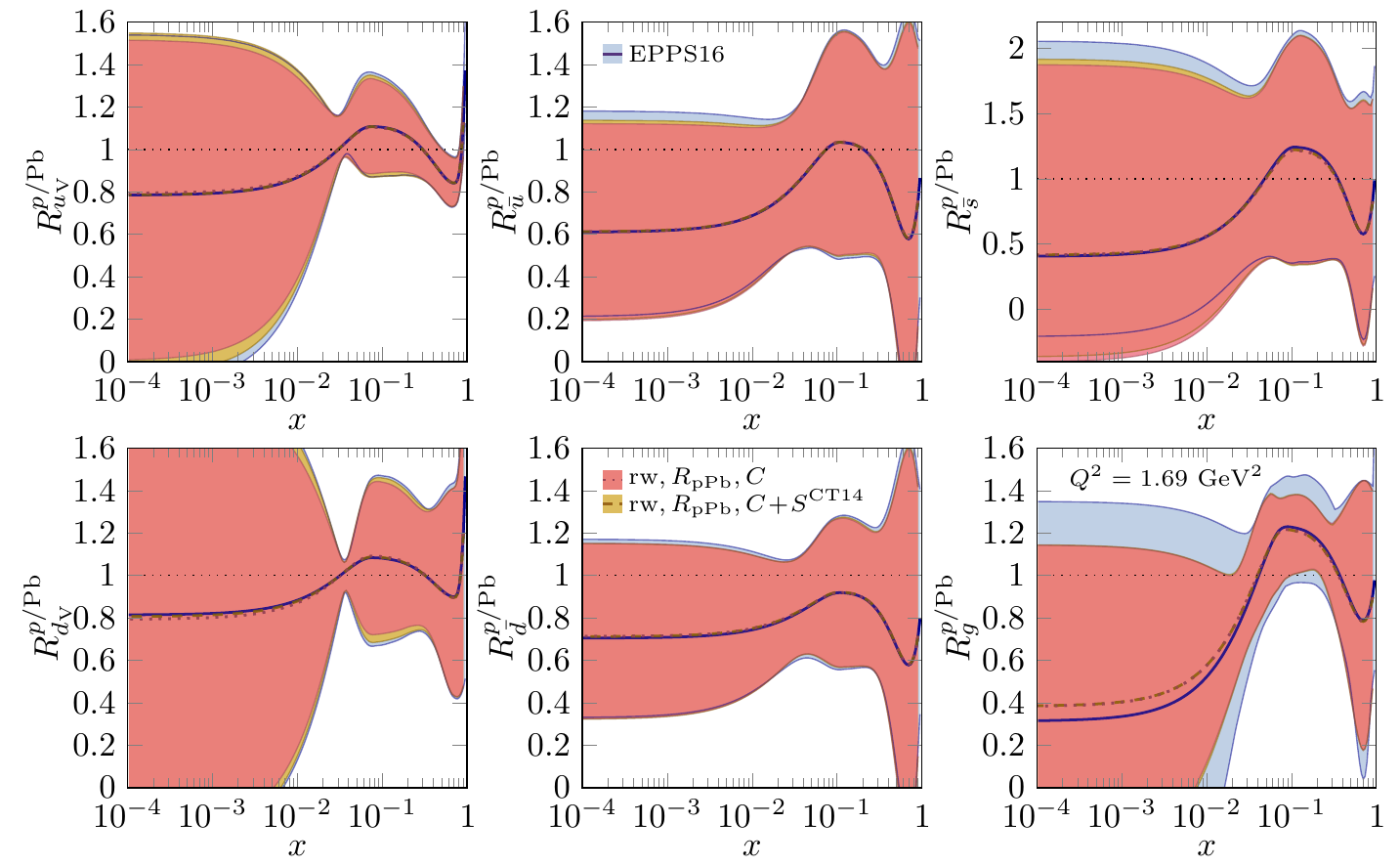}
  \caption{As Fig.~\ref{fig:rwwithabsxsec}, but now for reweighting with the nuclear-modification ratios from Fig.~\ref{fig:CMS_Ws_RpPb}.}
  \label{fig:rwwithnuclratio}
\end{figure*}

We have tested also the constraints obtainable from the different ratios discussed in Section~\ref{sec:unc_red}. The results are shown for the nuclear-modification ratios in Fig.~\ref{fig:rwwithnuclratio}, and we discuss the impact with the other ratios below. Overall, the constraints from the nuclear-modification ratios are even surprisingly similar to those with absolute cross sections. What we have gained by reducing the free-proton uncertainties, we have lost by introducing additional experimental uncertainties from the p+p reference. Importantly though, we do not lose any constraining power in the process. This could be improved further if one would be able to cancel some of the systematical uncertainties in the ratio. Still, using the nuclear-modification ratios reduces the difference in the reweighted uncertainties between the ``$C$'' and ``$C+S^\text{CT14}$'' extractions, the results being now almost identical. Also, while there is some pull on the central values of valence-quark nuclear modifications when using absolute cross sections, the size of which depends on whether the CT14 uncertainties are included, this effect vanishes with the nuclear modification ratio. Thus, with this observable and present data precision, the residual proton-PDF dependence in the nuclear-modification extraction is found to be small, conveniently for the global analyses. We observe also a slightly larger impact on the sea quarks compared to the absolute cross sections, but within the large uncertainties this is hardly significant. We emphasise that the nuclear-modification ratios were constructed here from separate measurements of p+Pb and p+p, and we therefore cannot cancel systematical uncertainties in the ratio, which could have improved the impact further. A direct measurement of the nuclear-modification ratios with the future LHC Run 3 data would thus be most welcome.

\begin{figure*}[tp]
  \centering
  \includegraphics[width=1.9\columnwidth]{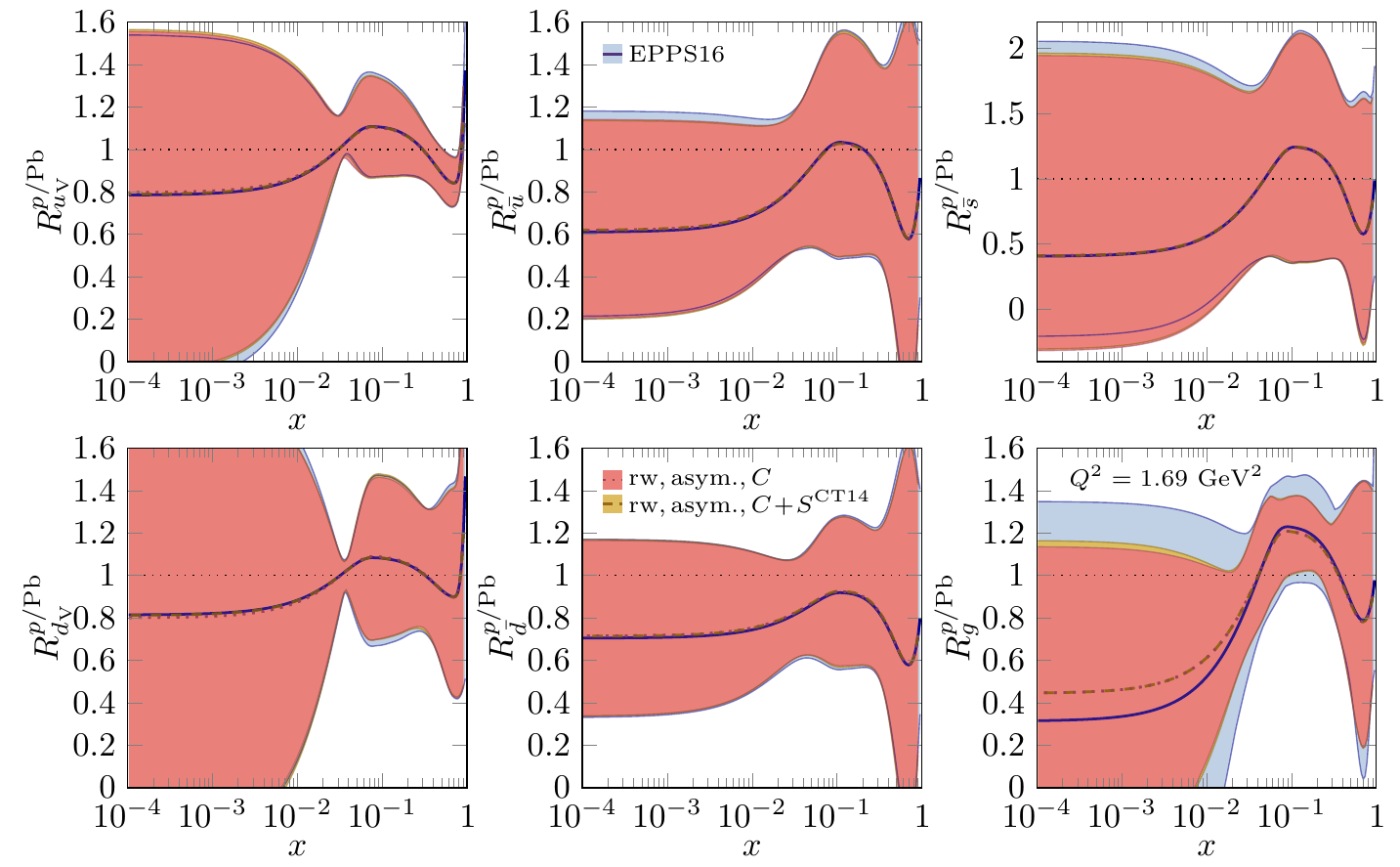}
  \caption{As Fig.~\ref{fig:rwwithabsxsec}, but now for reweighting with the alternative symmetries from Fig.~\ref{fig:CMS_Ws_other_asymmetries}.}
  \label{fig:rwwithasyms}
\end{figure*}

The relative smallness of free-proton-PDF impact is found to extend also to nuclear-modification extraction with the other ratios. For the self-normalized cross sections, where we fit by leaving one point out as explained in Section~\ref{sec:self-normalized}, the results are almost identical to those in Fig.~\ref{fig:rwwithnuclratio}, with only marginally larger proton-PDF dependence due to the less direct cancellation in the ratio. For the forward-to-backward ratios and charge asymmetries the loss of information restricts the achievable constraints from individual observables. As a result, with the forward-to-backward ratios essentially all constraints on the valence flavour separation are lost, and the resulting effect is on the more poorly constrained sea-quarks and gluons, exactly as in the middle and right panels of Fig.~\ref{fig:rwwithnuclratio}. As we have discussed above, the traditional charge asymmetry $\mathcal{A}_{\rm pPb}$ does not provide strong free-proton-PDF cancellation, and as a result the constraints on the valence quarks are the same as with the absolute cross sections in Fig.~\ref{fig:rwwithabsxsec}, but due to the cancellation of nuclear effects in the forward region, practically all constraints on the gluons are lost. When used separately, the alternative asymmetries $\tilde{\mathcal{A}}_{\rm pPb}$ and $\mathcal{A}^*_{\rm pPb}$ have essentially the same effect as the forward-to-backward ratios, with the $\mathcal{A}^*_{\rm pPb}$ being slightly more constraining than $\tilde{\mathcal{A}}_{\rm pPb}$, and constrain only the small-$x$ uncertainties which dominate in these observables. When used together, as shown in Fig.~\ref{fig:rwwithasyms}, these asymmetries begin to have some constraining power also on the valence quarks. The impact is small, but appears to be completely free from the proton-PDF uncertainties, as should be expected.

\begin{figure*}[t]
  \centering
  \includegraphics[width=1.9\columnwidth]{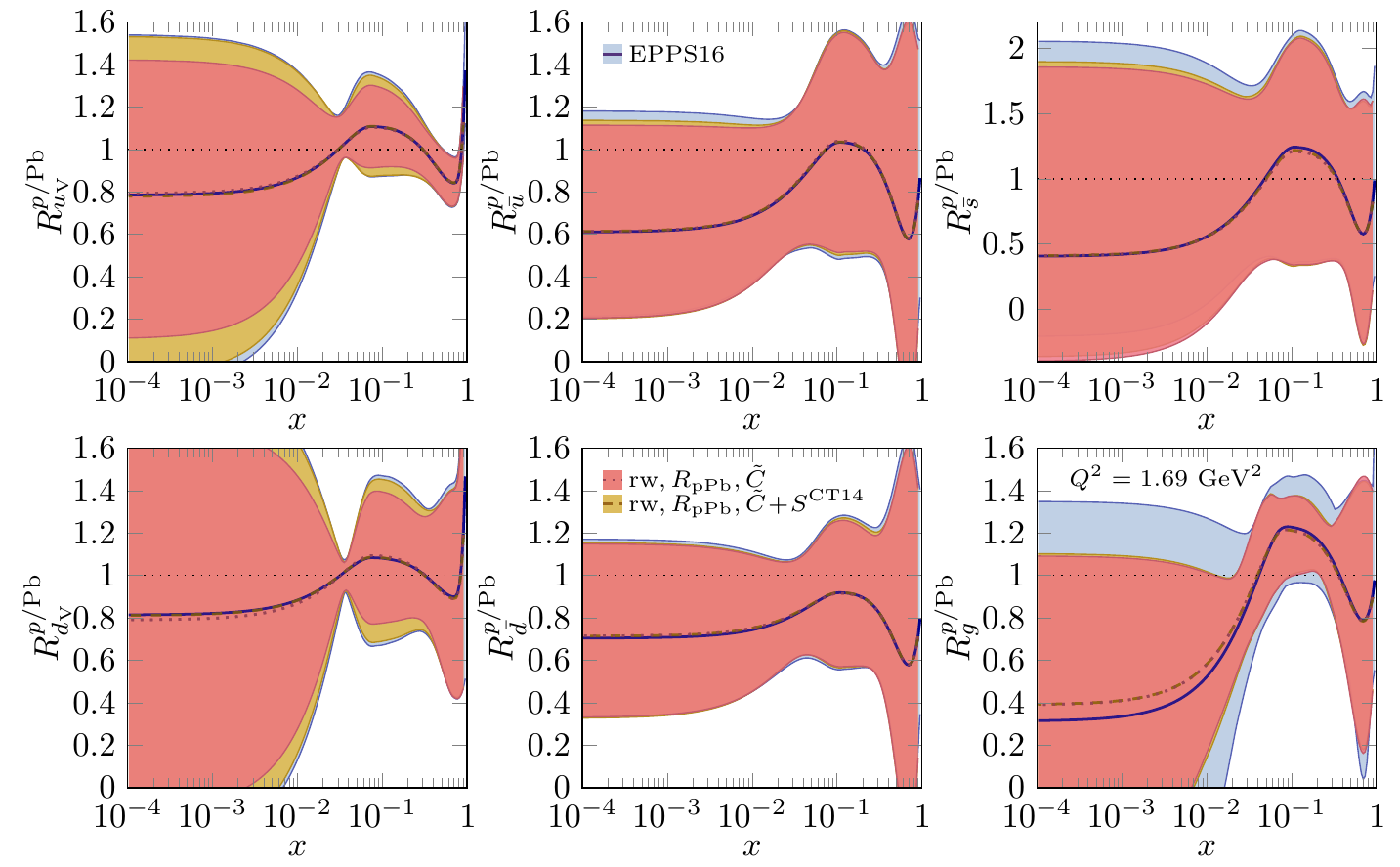}
  \caption{As Fig.~\ref{fig:rwwithnuclratio}, but using mock data with reduced statistical uncertainties.}
  \label{fig:rwwithreducedstatunc}
\end{figure*}

The impact on the valence-quark flavour separation appears to be limited mainly by the p+Pb statistical uncertainties. This will change after the LHC Run 3 data taking, where we expect a factor of 3--4 increase in the attainable statistics~\cite{Citron:2018lsq}. For a simple test, we have performed a reweighting with mock data with an altered covariance matrix $\tilde{C}$, where we take the current 8.16 TeV measurement and reduce the statistical uncertainties by a factor of two. As a result, the valence-quark nuclear-modification uncertainties become smaller when neglecting the free-proton uncertainties, but \emph{not} when the latter are taken into account. This holds even for relatively free-proton-PDF insensitive observables such as the nuclear-modification ratios, the results for which are shown in Fig.~\ref{fig:rwwithreducedstatunc}. As the $W^\pm$ boson measurements in p+p collisions are already dominated by systematical uncertainties (see Ref.~\cite{AbdulKhalek:2018rok} for discussion about the expected constraints from high-luminosity LHC) whereas in p+Pb there is still a significant statistical contribution to the uncertainty, it is plausible that the constraints on nuclear PDFs evolve faster than those on the proton PDFs (we found that changing CT14 to CT18 NLO PDFs did not have a significant effect on the free-proton uncertainties, but it should be noted that CT18 has relatively conservative uncertainty estimates compared to other contemporary analyses). The proton-PDF uncertainties can therefore in the worst case become even a limiting factor for extracting the bound-nucleon modifications in the future, and must then be propagated accordingly or suppressed with special ratios such as the alternative asymmetries discussed above.

\section{Summary and discussion}

In this paper, we have systematically studied the importance of the free-proton-PDF uncertainties in extracting the nuclear modifications of bound-nucleon PDFs from $W^\pm$ production data. We have done this in the context of the most recent lepton-rapidity-differential $W^\pm$ cross sections at 8.16 TeV~\cite{CMS:2019leu}, and various ratios constructed from the data. We have demonstrated that none of the considered ratios yield a perfect cancellation of the proton-PDFs uncertainties, with the exception of the alternative charge-asymmetry ratios discussed in Section~\ref{sec:asymmetries}. By using the methods of theoretical covariance matrices and Hessian PDF reweighting, we however find that the residual free-proton uncertainties in the ratios are small enough compared to the present data uncertainties that they do not pose significant bias in the nuclear-modification extraction.

The Hessian reweighting performed in this study also gives us information on what the impact of the 8.16 TeV $W^\pm$ production data would have been, had they been included in the EPPS16 analysis. We find that, after a consistent propagation or a reduction of the proton-PDF uncertainties, the impact on the flavour separation of the nuclear modifications is rather mild, owing to the existing constraints from CHORUS neutrino-DIS data~\cite{CHORUS:2005cpn} included in the EPPS16 analysis, and the main new information is on the nuclear gluons, which were poorly constrained in EPPS16. Importantly, the new gluon-PDF constraints are consistent with those~\cite{Eskola:2019bgf,Kusina:2017gkz,Eskola:2019dui} from dijet and D$^0$-meson production measurements~\cite{Sirunyan:2018qel,Aaij:2017gcy}, and the $W^\pm$ production data also fully agrees with the flavour-separation constraints from the CHORUS data. This supports the universality of nuclear PDFs. The simple reweighting study performed here falls short in few aspects which we address in a concurrent global analysis~\cite{Eskola:2021nhw} (see also the work in Refs.~\cite{AbdulKhalek:2019mzd,AbdulKhalek:2020yuc}): First, to reliably quantify the full free-proton-PDF dependence of the nuclear modifications, it is necessary to include them in all the relevant fitted observables. And second, once the free-proton uncertainties are included in the nuclear-modification fitting, one should find a way to consistently propagate these uncertainties into any desired observable (this problem is analogous to the one in evaluating scale uncertainties both in the PDF fits and predictions for observables, see Refs.~\cite{Harland-Lang:2018bxd} and~\cite{Ball:2021icz}).

Even though the impact of the free-proton PDFs was found here to be small in the nuclear-PDF extraction, especially when using the experimental nuclear-modification ratios, this can radically change in the future with yet another 3--4 fold increase in the p+Pb statistics expected from the LHC Run 3. As argued in Section~\ref{sec:rw}, the proton-PDF uncertainties can then become a significant source of uncertainty in the extraction of the nuclear modifications, which highlights the importance of understanding the large-$x$ nucleon structure~\cite{Accardi:2011fa,Dudek:2012vr} and accurate determination of the free-proton PDFs and their uncertainties also from the nuclear PDF point of view. If we no longer can take the (effective) bound-nucleon nuclear modifications to be independent of the free-nucleon PDFs themselves, this can have also significant consequences e.g.\ for the attempts to understand the physical cause behind the EMC effect~\cite{Arrington:2019wky}. In this case, special observables like the alternative asymmetries can prove to be useful in testing different models.

We stress that this study (in its present extent) would not have been possible without having access to the experimental data correlations. When we strive for an increased precision in the nuclear-PDF extraction with the upcoming LHC runs and future experiments, it becomes increasingly important to publish the full covariance matrices of the measurements or, even better, a full breakdown of the systematical-uncertainty contributions, source by source, optimally with the full information on the used statistical model published as well~\cite{Cranmer:2021urp}. This would enable nuclear-PDF fits to use the maximal amount of information, with the possibility to even ``cross calibrate'' different measurements~\cite{Helenius:2019lop}.

\section*{Acknowledgments}

We thank Francois Arleo for discussions concerning the asymmetry observables in $W^\pm$ production. We acknowledge financial support from Xunta de Galicia (Centro singular de investigación de Galicia accreditation 2019-2022), European Union ERDF, the “María de Maeztu” Units of Excellence program MDM-2016-0692 and the Spanish Research State Agency, European Research Council project ERC-2018-ADG-835105 YoctoLHC (C.A.S. and P.P.), and from the Academy of Finland projects 308301 (H.P.) and 330448 (K.J.E. and P.P.). This research was funded as a part of the Center of Excellence in Quark Matter of the Academy of Finland (project 346325).

\bibliographystyle{spphys}
\bibliography{article_CMS_Ws_EPJC}

\end{document}